\documentclass[useAMS,usenatbib]{mn2e}
\usepackage{psfrag}
\usepackage{graphicx}
\usepackage{color}
\usepackage{epsfig}
\usepackage{amsmath,amssymb}
\usepackage{natbib}
\usepackage[abs]{overpic}
\usepackage{pict2e}

\def\diff{\mathrm{d}}
\def\ddt#1#2{\frac{\diff #1}{\diff #2}}
\def\pddt#1#2{\frac{\partial #1}{\partial #2}}

\def\simge{
    \mathrel{\rlap{\raise 0.511ex
        \hbox{$>$}}{\lower 0.511ex \hbox{$\sim$}}}}
\def\simle{
    \mathrel{\rlap{\raise 0.511ex
        \hbox{$<$}}{\lower 0.511ex \hbox{$\sim$}}}}

\newcommand{\figref}[1]{Figure~\ref{#1}}
\newcommand{\tabref}[1]{Table~\ref{#1}}
\newcommand{\secref}[1]{Section~\ref{#1}}

\newcommand{\msun}{M_{\odot}}
\newcommand{\rsun}{R_{\odot}}

\newcommand{\msunyr}{M_\odot~{\rm yr}^{-1}}

\newcommand{\hii}{H{\sc ii} }

\title[Formation of primordial supermassive stars by burst accretion]
{Formation of primordial supermassive stars by burst accretion}
\author[Y. Sakurai, T. Hosokawa, N. Yoshida, and H. W. Yorke]{Y. Sakurai$^{1}$, 
T. Hosokawa$^{2}$, N. Yoshida$^{3}$, and H. W. Yorke$^{4}$
\\
$^{1}$Department of Physics, The University of Tokyo, Tokyo 113-0033, Japan\\
$^{2}$Department of Physics and Research Center for the Early Universe, The University of Tokyo, Tokyo 113-0033, Japan\\
$^{3}$Kavli Institute for the Physics and Mathematics of the Universe (WPI), The University of Tokyo, Kashiwa, Chiba 277-8583, Japan\\
$^{4}$Jet Propulsion Laboratory, California Institute of Technology, Pasadena CA 91109, USA
}
\begin{document}

\date{Draft version \today}


\maketitle

\label{firstpage}

\begin{abstract}
Recent observations show that supermassive black holes 
(SMBHs) with $\sim10^9~M_\odot$ exist at redshift $z\gtrsim6$;
how they form has yet to be explained.
A promising formation channel is the so-called 
direct collapse model, 
which posits that a massive seed BH forms through
gravitational collapse of a $\sim 10^5~M_\odot$ supermassive star.
We study the evolution of such a supermassive
star growing by rapid mass accretion. The internal
stellar structure is also followed consistently in 
our calculation.
In particular, we examine the impact of time-dependent
mass accretion of repeating burst
and quiescent phases that are expected to occur
with a self-gravitating circumstellar disk. 
We show that the stellar evolution with such
episodic accretion differs qualitatively from that expected with 
a constant accretion rate, even if the mean accretion 
rate is the same. Unlike the case of constant mass accretion, 
whereby the star expands roughly following 
$R_* \simeq  2.6 \times 10^3 \rsun (M_*/100~\msun)^{1/2}$,
the protostar can substantially contract during the quiescent
phases between accretion bursts.
The stellar effective temperature and ionizing photon emissivity 
increase accordingly as the star contracts, which can cause strong
ionizing feedback and halt the mass accretion onto the star.
With a fixed duration of the quiescent phase $\Delta t_{\rm q}$, 
such contraction occurs in early evolutionary phases, i.e.
for $M_* \lesssim 10^3~\msun$ with 
$\Delta t_{\rm q} \simeq 10^3$ yr. For later epochs and 
larger masses but the same $\Delta t_{\rm q}$, 
contraction is negligible even during quiescent phases.
With larger quiescent times $\Delta t_{\rm q}$, however, 
the star continues to contract during quiescent phases even for the
higher stellar masses.
We show that such behavior is well understood by comparing the
interval time and the thermal relaxation
time for a bloated surface layer. 
We conclude that the UV radiative feedback becomes
effective if the quiescent phase associated by the
burst accretion is longer than $\sim 10^3$ yr, which is possible
in an accretion disk forming in the direct collapse model.
\end{abstract}


\section{Introduction}
\label{sec:Introduction}

Recent observations have revealed the existence of supermassive black
holes (SMBHs) with masses of $\sim 10^9~M_\odot$ at $z \gtrsim 6$
~\citep[e.g., ][]{2015Natur.518..512W, Marziani:2012rf, 2011Natur.474..616M}.
The origin and the rapid growth of the early SMBHs remain to be elucidated.


Possible seeds of the SMBHs are remnant BHs of Population III stars
with $M_{\rm BH} \sim 100~M_\odot$
\citep[e.g., ][]{Madau:2001lr,Schneider:2002lr}.
Although such a $\sim100~M_\odot$ seed BH can just barely attain a mass of
$\sim 10^9~M_\odot$ by the epoch of $z \simeq 6$,
by accreting material at the Eddington rate, 
recent studies suggest difficulties in this model.
For instance, radiative feedback from a BH accretion disk 
easily suppresses the gas supply
from the intergalactic medium, so that the accretion rates
fall far below the Eddington values 
\citep[e.g., ][]{Alvarez09,Jeon:2012qy}.
The BH growth time then becomes much longer than the age of the
universe at  $z \sim 6$,


An alternative model which circumvents these difficulties is the so-called
direct collapse model.
The model assumes that SMBHs are built from larger seed BHs 
with $M_{\rm BH} \sim 10^{5}~M_\odot$ that are formed directly 
by gravitational collapse of supermassive stars (SMSs) 
of similar masses \citep[e.g., ][]{Bromm:2003uq}.
The BH growth time is significantly shortened in this case; 
a $\sim10^{5}~M_\odot$ seed BH can easily grow to become 
a $\sim10^{9}~M_\odot$ SMBH 
by accretion at the Eddington rate.
At reduced accretion rates, e.g. due to
radiative feedback from a BH-disk system, a
$\sim 10^9~M_\odot$
SMBH can still form at $z\simeq6$
if the mean accretion rate is higher than
about 50~\% of $\dot{M}_\mathrm{Edd}$.
Indeed, cosmological simulations show that this is possible
with efficient cold accretion flows realized in the epoch
of the first-galaxy formation \citep[e.g.,][]{DiMatteo12}.


A crucial assumption in the direct collapse model is
that sufficiently massive stars are formed in a gas cloud.
SMSs are supposed to form in so-called atomic-cooling
haloes with $T_{\rm vir} \sim 10^4$~K\citep[e.g.,][]{IO12,Agarwal14,Visbal14}.
If molecular hydrogen cooling is inhibited in these massive 
primordial haloes by strong ultra-violet radiation,
a gas cloud gravitationally collapses 
via atomic hydrogen cooling nearly isothermally 
at $T \simeq 8000$~K \citep[e.g.,][]{Omukai01}.
An embryo protostar eventually forms and begins to grow
via gas accretion from a surrounding envelope
\citep[e.g.,][]{IOT14,VanBorm14},
analogous to normal Population III star formation \citep[][]{YOH08}. 
The accretion rate at this stage has the well-known temperature dependence
\begin{equation}
\dot{M_*}\sim \frac{c_\mathrm{s}^{3/2}}{G}\sim 0.2
\left(\frac{T_\mathrm{vir}}{10^4}\right)^{3/2}~\msunyr,
\label{eq:accretion rate}
\end{equation}
where $c_\mathrm{s}$ is the sound speed and 
$G$ is the gravitational constant.
Because of the higher gas temperature during the collapse stage, the
accretion rate is much higher than that for normal Pop III cases.
If such rapid mass accretion continues for $\sim 1~{\rm Myr}$,
which is the lifetime of massive stars, 
SMSs with $10^5 - 10^6~M_\odot$ can form and
ultimately provide massive BH seeds after their deaths.


A possible obstacle for the growth of SMSs via rapid mass accretion
is strong radiative feedback from the central massive protostar.
Even when a normal Pop III star with $\sim100~M_\odot$ forms
by accretion, stellar UV radiation becomes so strong to 
create an \hii region that dynamically expands through 
the accretion envelope\citep[e.g.][]{McKee:2008fj}.
The gas accretion is then halted; the final stellar mass
is effectively determined by this mechanism
\citep[]{HOYY11,Hirano14}.
For the case of SMS formation considered here, 
similar, or even stronger UV feedback might significantly 
reduce or even halt the gas mass accretion.
This could occur because a SMS should emit copious amounts of
photons; the stellar luminosity is nearly at
the Eddington value, which for massive stars is proportional to stellar mass.
The resulting UV feedback could thus be much stronger than
for normal Pop III cases. 


The strength of the stellar radiative feedback depends 
critically on the
ionizing photon emissivity of primordial protostars,
which is determined by their evolution while accreting.
For normal Pop III cases with mean accretion rates
of a few $\times 10^{-3}~\msunyr$, an accreting protostar
has a large radius during the early evolution but eventually 
contracts to the zero-age main-sequence (ZAMS) 
for $M_* \sim 100~\msun$ \citep[e.g.,][]{Omukai:2003kq}. 
When the protostar contracts, its ionizing photon emissivity
increases and UV radiative feedback becomes effective.


By contrast, the protostellar evolution should qualitatively
differ from the above at very rapid accretion rates
$\dot{M_*} \gtrsim \dot{M}_{\rm *,cr} = 4 \times 10^{-2}~\msunyr$
expected in the direct collapse model. 
In this case, a protostar continues to expand and does not contract 
even after the stellar mass greatly exceeds $100~\msun$ \citep[][]{Hosokawa:2012lr}. 
The stellar radius reaches
$\sim 10^4~\rsun$ for $M_* \gtrsim 10^3~\msun$ 
\citep[][]{Hosokawa:2013jk,Schleicher13}.
Such a ``supergiant protostar'' has a low effective temperature
of $T_{\rm eff} \simeq 5000$~K and the ionizing photon emissivity for
$M_* \lesssim 10^4~M_\odot$ is only
$\lesssim10^{45} \mathrm{~s}^{-1}$, several
orders of magnitude lower than for main-sequence SMSs.
The resulting UV feedback should be too weak to halt
the mass accretion. 
The rapid mass accretion should thus continue, so that 
SMSs finally form.



The previous studies consider 
only constant accretion rates.
In more realistic situations, however, the mass accretion onto 
a growing SMS should be dynamic with highly time-dependent 
accretion histories. For normal Pop III cases, for instance, 
an accretion disk formed around a protostar becomes
gravitationally unstable and fragments 
\citep[e.g.,][]{Stacy10,Greif11}. 
When such fragments migrate inward in the disk and accrete
onto the star \citep[e.g.,][]{Greif12}, the accretion rate 
drastically increases causing so-called ``burst accretion''
\citep[e.g.,][]{Vorobyov:2013lr}.
In this case, the accretion history is roughly divided 
into two phases: the burst phases with
$\dot{M_*} \sim 0.1~\msunyr$ and
quiescent phases with $\dot{M_*} \sim 10^{-4}~\msunyr$,
whose durations are $\lesssim 10^2 \mathrm{~yr}$ and 
$\sim 10^3-10^4 \mathrm{~yr}$.
Such episodic accretion is also expected to occur in the direct
collapse model~\citep[e.g.,][]{Inayoshi:2014qy}.
Indeed, recent high-resolution numerical simulations report
signatures of the disk fragmentation in atomic-cooling
haloes \citep[][]{Regan:2014rf,Becerra14}. 


In this paper, we study the evolution of SMSs
with time-dependent accretion histories in which a 
number of accretion bursts occur.
Whereas the radiative feedback is generally weak for
constant accretion, a time-dependent accretion
rate might significantly affect the protostellar evolution
and hence the feedback strength. 
In quiescent phases between the bursts, for instance,
accretion rates should fall below the critical rate
$\dot{M}_{\rm *,cr} = 4 \times 10^{-2}~\msunyr$, above which
a protostar enters the supergiant stage. 
The protostar may contract in this case, so that the effective 
temperature rises and radiative feedback becomes effective.
Here, we show that, by solving the stellar interior structure numerically,
this should occur even with a mean accretion rate of 
$\sim 0.1~\msunyr$, if the quiescent phase lasts for 
periods $\gtrsim 10^3$ yr.


The remainder of this paper is organized as follows.
In \secref{sec:Numerical method} our numerical
methods and modeling of the burst accretion are explained.
The numerical results are presented in \secref{sec:Results},
where we also compare the protostellar evolution for
constant accretion rates to the evolution for time-dependent rates with
a number of accretion bursts.
We discuss implications of these results in \secref{sec:discussion} and
summarize our conclusions in \secref{sec:Conclusions}.

\section{Numerical method and modeling of burst accretion}
\label{sec:Numerical method}

\subsection{Numerical method}


To calculate the stellar evolution, we use the numerical code
originally developed by \citet{Yorke:2008rz} with additional
improvements as in \citet{Hosokawa:2013jk}.
The code solves the following four basic equations of the stellar evolution:
\begin{equation}
\pddt{r}{m}=\frac{1}{4\pi r^2 \rho}~,
\label{eq:Mconv}
\end{equation} 
\begin{equation}
\pddt{P}{m}=-\frac{Gm}{4\pi r^4}~,
\label{eq:Euler}
\end{equation}
\begin{equation}
\pddt{l}{m}
=E_\mathrm{nuc}
-c_P\pddt{T}{t}+\frac{\delta}{\rho}\pddt{P}{t} 
\label{eq:EnergyConservation}
\end{equation}
\begin{equation}
\pddt{T}{m}
=-\frac{GmT}{4\pi r^4 P}\nabla,
\label{eq:EnergyTransport}
\end{equation}
where $m$ is the mass coordinate, $r$ the radial distance from the center, 
$P$ the total (radiation plus gas) pressure,
$l$ the local luminosity, $T$ the temperature, $\rho$ the density, 
$E_\mathrm{nuc}$ the net energy generation rate by 
nuclear fusion, $c_P$ the isobaric specific heat, 
$\nabla \equiv \partial\ln T/\partial\ln P$ the actual temperature 
gradient, and 
$\delta \equiv -\left(\partial\ln\rho/\partial\ln T\right)_P$.
The temperature gradient $\nabla$ is evaluated in the context of
mixing-length theory for convective layers.
The nuclear reactions of hydrogen and helium burning are included
in the code. 


In Eq. \eqref{eq:Euler}, hydrostatic equilibrium is always 
assumed and the inertial term is omitted.
This assumption is well-founded, because 
the local free-fall time in the stellar interior 
is generally much shorter than 10 years,
the typical timescale over which
the mass accretion rate varies
(also see Sec.~\ref{subsec:Modeling of burst accretion} below). 
Note that we consider the time-derivative in Eq. 
\eqref{eq:EnergyConservation}. Thus, thermal equilibrium is
{\it not} assumed in our calculations.


Mass accretion is implemented by adding 
a mass $\dot{M_*}\Delta t$ at each time step to the 
outermost layer of stellar models.
We assume that the newly accreting gas has the same
physical quantities as in the stellar atmosphere.
This will be realized if the accreting materials 
have enough time to adjust their thermal states to 
those in the atmosphere, e.g., 
when slowly approaching the stellar surface orbiting
in a circumstellar disk.
This is obviously a limiting case, and the accreting
gas may have some additional thermal energy. 
We model this potential difference using a free-parameter 
$\eta$, which is the fraction of the 
energy advected into the star with the accreting gas
to the released gravitational energy, 
\begin{equation}
\eta \equiv \frac{L_{*,\mathrm{acc}}}{L_{\rm acc}}
= L_{*,\mathrm{acc}}  
  \left( \frac{G M_* \dot{M}_*}{R_*} \right)^{-1}
~,
\end{equation}
where $L_{*,\mathrm{acc}}$ is the additional energy input 
to the star.
The values of $\eta$ adopted in our set of models are summarized in 
\tabref{tab:accretion} 
(also see \secref{subsec:Modeling of burst accretion}).
As explained in \citet{Hosokawa:2013jk}, however, $\eta$ actually has
little effect on the stellar evolution except during early
stages when $M_* \lesssim 100~M_\odot$.


We follow the growth of a primordial supermassive protostar by numerically
solving the stellar structure equations.
We consider several models of accreting protostars that grow
via episodic accretion (see \ref{subsec:Modeling of burst accretion} below).
Our calculations begin with an initial pre-main sequence stellar model of 
mass $2~M_\odot$ and radius $25~R_\odot$, whereby the initial
interior structure is determined by solving the Lane-Emden 
equation with a polytropic index $n=1.5$.
We assume that the initial model and newly accreting gas
have the primordial composition with $X=0.72$ and 
$Y=0.28$.

\subsection{Modeling of burst accretion}
\label{subsec:Modeling of burst accretion}

\begin{table*}
 \begin{center}
  \caption{Models of episodic accretion whose mean accretion rates are about 
 $0.1~M_\odot$/yr.}
   \begin{tabular}{|l|rrrr|} \hline
     Model & A & B & C & D \\ \hline
     Duration of the burst phase $\Delta t_{\rm b}$ [ yr ] & 25  & 50 & 100 & 500 \\
     Duration of the quiescent phase $\Delta t_{\rm q}$ [ yr ] & 270  & 540 & 1080 & 5400 \\ \hline
     Accretion rate in the burst phase $\dot{M}_{\rm *,b}$  [ $M_\odot/\mathrm{yr}$ ] & 1  & 1  & 1 & 1 \\ 
     Accretion rate in the quiescent phase $\dot{M}_{\rm *,q}$ [ $M_\odot/\mathrm{yr}$ ] & $10^{-3}$ & $10^{-3}$ & $10^{-3}$ & $10^{-3}$ \\ \hline
     Transition time $\Delta t_{\rm t}$ [ yr ] & 50 & 100 & 200 & 1000 \\ \hline
     Radiative efficiency $\eta$ & 0.1 & 0.1 & 0.1 & 0.01 \\ \hline
   \end{tabular}
  \label{tab:accretion}
 \end{center}
\end{table*}


We model accretion histories expected in the case of a self-gravitating 
accretion disk. Numerical simulations show that, in such cases, 
the accretion rate sometimes sharply increases like a burst. 
For instance, such a burst occurs when a fragment formed via the 
gravitational instability migrates inward through the disk to 
be accreted onto the star \citep{Vorobyov:2013lr}.  
Normally, such a burst is followed by a quiescent phase,
when the mass accretion almost ceases. 
Another accretion burst can be triggered when the disk 
self-gravity is effective again due to mass growth of the disk
from a surrounding envelope. 
We study the effect of burst accretion for the case of 
the direct collapse model.


We model accretion histories with
periodic accretion bursts \citep[e.g.,][]{BCG09,HOK11}, by
assuming alternating constant high and low accretion rates
for the burst and quiescent phases, respectively:
$\dot{M}_{\rm *,b}$ and $\dot{M}_{\rm *,q}$.
The durations of these phases are parametrized with
$\Delta t_{\rm b}$ and $\Delta t_{\rm q}$.
We also assume that the transition from a burst 
to a quiescent phase takes place over a finite duration
$\Delta t_{\rm t}$.
The accretion rates during the transition phases are given
by linear interpolation.
In total, we have five parameters that characterize a model
accretion history: $\dot{M}_{\rm *,b}$, $\dot{M}_{\rm *,q}$,
$\Delta t_{\rm b}$, $\Delta t_{\rm q}$, and $\Delta t_{\rm t}$.


We examine four different cases with different sets of these
parameters as listed in \tabref{tab:accretion}.
The values are chosen so that the mean accretion rate 
is $0.1~M_\odot/\mathrm{yr}$, which 
is expected for the direct collapse model.
We compare the results of the four cases with effectively
the same mean accretion rate.
We start the evolutionary calculations with
an accretion burst.


The finite transition phase is needed for numerical stability when
calculating stellar evolution with highly 
time-dependent mass accretion histories. 
We assume finite
transition phases with 
$\Delta t_{\rm b} \lesssim \Delta t_{\rm t} < \Delta t_{\rm q}$ 
(see \tabref{tab:accretion} for the adopted values).
We have checked that varying the transition time by a factor of a few
has little effect on the results.

\section{Results}
\label{sec:Results}
\subsection{Evolution with constant accretion rates}

\subsubsection{Normal Pop III case with $10^{-3}~\msunyr$}
\label{sec:0.001}

Before presenting the protostellar evolution with the burst accretion,
we briefly describe the fiducial case 
with a constant accretion rate for comparison. 
In the typical case with the accretion rate of $10^{-3}\msunyr$,
a protostar goes through several distinct evolutionary stages. 
Let us consider the following two timescales 
\citep[e.g.,][]{SPS86,Omukai:2003kq,Hosokawa:2009rf}: 
the Kelvin-Helmholtz (KH) time
\begin{equation}
t_\mathrm{KH}=\frac{GM_*^2}{R_*L_*},
\label{eq:KH}
\end{equation}
which is the thermal relaxation time over which the gravitational 
energy is released by radiation, and the accretion time
\begin{equation}
t_\mathrm{acc}=\frac{M_*}{\dot{M_*}},
\label{eq:acc}
\end{equation}
which is the characteristic stellar growth time. 
In the earliest phase when $M_* \lesssim 5~\msun$, the star evolves
adiabatically, satisfying $t_{\rm acc} < t_{\rm KH}$.
As the stellar luminosity $L_*$ increases 
with increasing stellar mass, the KH time becomes shorter.
The luminosity increases because the 
the stellar interior opacity decreases according to
Kramers' law $(\kappa \propto \rho T^{-3.5})$
when the temperature increases.
As the opacity decreases, the heat accumulated in the interior gradually 
escapes outward by radiative heat transport.
The timescale balance is finally inverted and the protostar
begins to contract radiating the energy away from the surface.
The KH contraction stage begins when $M_* \gtrsim 8~\msun$
in the case of $\dot{M} = 10^{-3}~\msunyr$, 
as can be seen in Figure~\ref{fig:MDOTMR}.
The stellar interior temperature rises during the KH contraction
stage, and finally hydrogen burning begins at the
center when $M_* \simeq 40~\msun$.
After this point, the evolution of the stellar radius closely traces 
the mass-radius relation of a ZAMS star. 


Figure~\ref{fig:MDOTMR} shows that the rapid rise of the stellar
ionizing photon emissivity is synchronous with the KH contraction.
Before KH contraction begins, the ionizing photon emissivity is only 
$\sim10^{37}~\mathrm{sec}^{-1}$. 
As the star contracts and the effective temperature rises, however,
the emissivity quickly reaches $\gtrsim 10^{49}~\mathrm{sec}^{-1}$.
Radiation hydrodynamic simulations show that the stellar radiation feedback 
operates in the late KH stage and eventually shuts off 
the mass accretion onto the star \citep[e.g.,][]{HOYY11}.

\subsubsection{Direct collapse case with $0.1~\msunyr$}
\label{sec:0.1}

\begin{figure}
\centering
\resizebox{79mm}{!}{\includegraphics{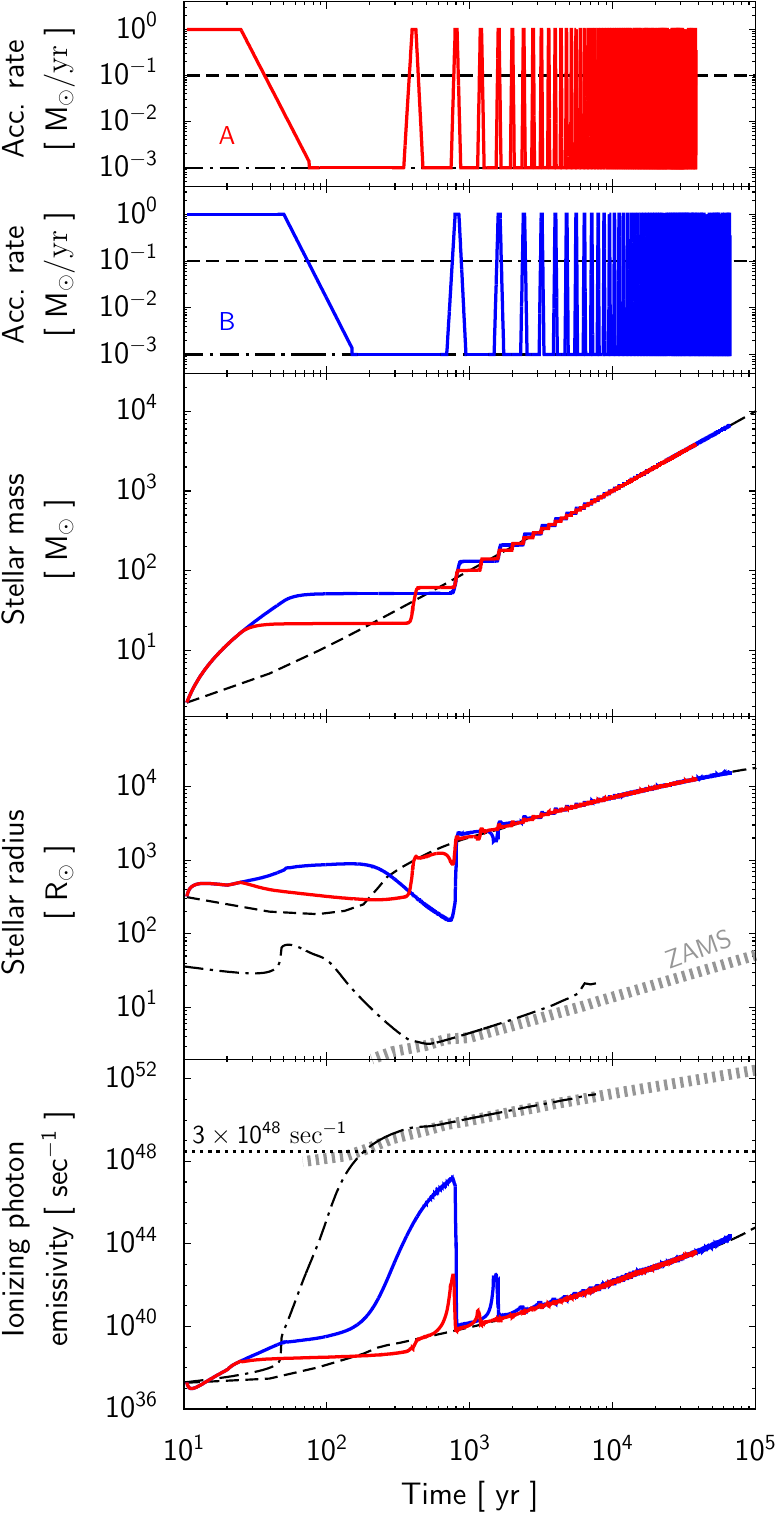}}
 \caption{Time evolution of the
accretion rates (top two panels),
 stellar mass (third panel), 
 radius (fourth panel), and ionizing photon 
 emissivity (bottom panel).
 In each panel, the red and blue lines represent Models A and B 
 (see \tabref{tab:accretion}).
 The evolution with constant accretion rates of $0.1~\msunyr$
 and $10^{-3}~\msunyr$ are also presented with the black dashed 
 and dotted-dashed lines respectively.
 In the bottom two panels, 
 the lines for $10^{-3}~\msunyr$ have been horizontally shifted so that
 the increase of stellar mass matches that for 
 $0.1~\msunyr$, i.e., $M_* = 100~\msun$ at
 the time of $10^3$ years. 
 The radii and ionizing photon emissivity of ZAMS stars are also plotted with
 the thick gray dashed lines, which are also plotted assuming the same
 mass variation. In the bottom panel the horizontal short-dashed line 
 represents the critical ionizing photon emissivity above which UV 
 feedback should become effective with an accretion rate  
 $0.1~\msunyr$ (see \secref{sec:discussion}).}
 \label{fig:MDOTMR}
\end{figure}

Here, we consider protostellar evolution at the high
constant accretion rate of $0.1~\msunyr$.
Unlike the case for $10^{-3}~\msunyr$,  
the protostar does not evolve to the KH contraction stage 
(see the dashed lines in Fig.~\ref{fig:MDOTMR}).
The star instead continues to expand 
nearly monotonically with increasing mass.

\begin{figure}
 \begin{center}
\resizebox{80mm}{!}{\includegraphics{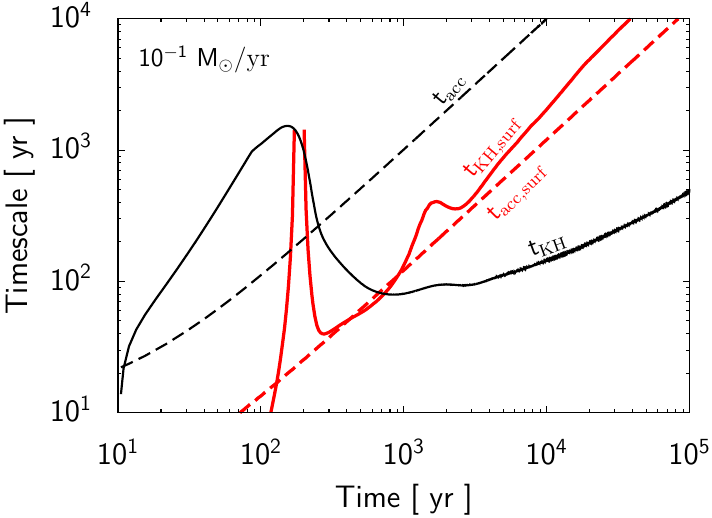}}
 \caption{
 The evolution of timescales with the constant accretion rate of
 $0.1~\msunyr$. The black solid and dashed lines represent KH timescale
 $t_\mathrm{KH}$ and accretion timescale $t_\mathrm{acc}$, respectively.
 The red solid and dashed lines are the local KH and accretion timescales
 $t_\mathrm{KH,surf}$ and $t_\mathrm{acc,surf}$.
 }
 \label{fig:KHtime-1}
 \end{center}
\end{figure}

The timescale inversion described in Section~\ref{sec:0.001} 
occurs at $M_* \simeq 22~\msun$ ($t \simeq 200$ yr)
in this case (Fig.~\ref{fig:KHtime-1}). 
The star evolves adiabatically before this epoch. 
Since the rapid accretion enhances the average entropy 
in the stellar interior,
the stellar radius 
at a given mass is larger than for $10^{-3}~\msunyr$.


The protostar continues expanding even when
$t_\mathrm{KH} < t_\mathrm{acc}$;
the star radiates the internal energy through the surface, 
which normally makes the star contract.
\citet{Hosokawa:2012lr} show that most of the stellar
interior actually contracts even in this bloating phase.
Basically, only the newly accreted surface layer 
inflates by absorbing the outward heat flux coming from the 
contracting interior. The mass of the bloating surface layer 
has a small fraction of the total stellar mass.
In the surface layer, the opacity is mostly due to
 H$^-$ bound-free
absorption that has a very strong temperature-dependence.
As a result, the stellar effective temperature is locked at 
a constant value $\simeq 5000$~K as in the case of red-giants.
From relations $T_{\rm eff} \simeq 5000$~K and 
$L_* = 4 \pi R_*^2 T_{\rm eff}^4 \simeq L_{\rm Edd}$, we can
easily show 
\begin{equation}
R_* \simeq 2.6 \times 10^3~R_\odot
\left( \frac{M_*}{100~\msun} \right)^{1/2},
\label{eq:ranalytic}
\end{equation}
which explains the numerical results very well.
The evolution of ionizing photon emissivity also differs 
from the case with $\dot{M} = 10^{-3}~\msunyr$.
Even when $M_* > 100~\msun$, the protostar still has very low 
ionizing photon emissivity
because of the low effective temperature. 
The resulting stellar UV feedback is then too weak
to halt the rapid mass accretion~(see \secref{sec:discussion}). 


We have found that in spite of
the timescale imbalance $t_\mathrm{KH} < t_\mathrm{acc}$,
the star continues to expand.
The apparent discrepancy is explained by the fact that the normal definition of
$t_\mathrm{KH}$ (Eq. \ref{eq:KH}) is a global relationship and does not consider the stellar
interior structure; as we discussed above, 
most of the stellar mass is
concentrated near the center and 
the bloated surface layer contains only a small fraction of the stellar mass
~\citep[also see Figure~2 in][]{Hosokawa:2013jk}.


We therefore must evaluate the local KH timescale for the surface layer
by considering the actual mass distribution within a star,
\begin{equation}
t_\mathrm{KH,surf} = 
\frac{f\int s_\mathrm{rad}T\mathrm{d}m}{\int \mathrm{d}l} ,
\label{eq:KHsurf}
\end{equation}
where $s_\mathrm{rad}$ is the entropy of radiation and $f$ is 
a dimensionless constant of $\mathcal{O}(1)$. 
The latter factor is introduced to represent that 
only a fraction $f$ of the total entropy is
carried away over the timescale.
We compare the above local KH time with the local 
accretion timescale for the surface layer, 
\begin{equation}
t_\mathrm{acc,surf}=\frac{f\int\mathrm{d}m}{\dot{M_*}}.
\label{eq:accsurf}
\end{equation}
For consistency we also include the same factor $f$ in 
Eq. \eqref{eq:accsurf}.
In the following, we set $f = 0.4$ as a fit to
our numerical results. 
We take the integration range 
in Eqs. \eqref{eq:KHsurf} and \eqref{eq:accsurf} 
as $0.7~M_* \leq m \leq M_*$ to cover the surface layer. 
We have checked that varying the lower bound does not 
significantly affect the main results.
\figref{fig:KHtime-1} shows that the local accretion time 
$t_{\rm acc, surf}$ still remains shorter than the corresponding 
KH time $t_{\rm KH, surf}$ even when the global KH timescale
satisfies $t_{\rm acc} > t_{\rm KH}$. 
The star continues to expand because,
with the rapid mass accretion, the gas accumulates in the surface 
layer faster than it can thermally relax. 
In other words, part of the outgoing heat flux is trapped by the accreting
gas, and the inflated surface layer cannot contract.

Below we show that these two local timescales
$t_{\rm acc, surf}$ and $t_{\rm KH, surf}$ are also useful for understanding
stellar evolution for the case of burst accretion.


\begin{figure}
\centering
\resizebox{79mm}{!}{\includegraphics{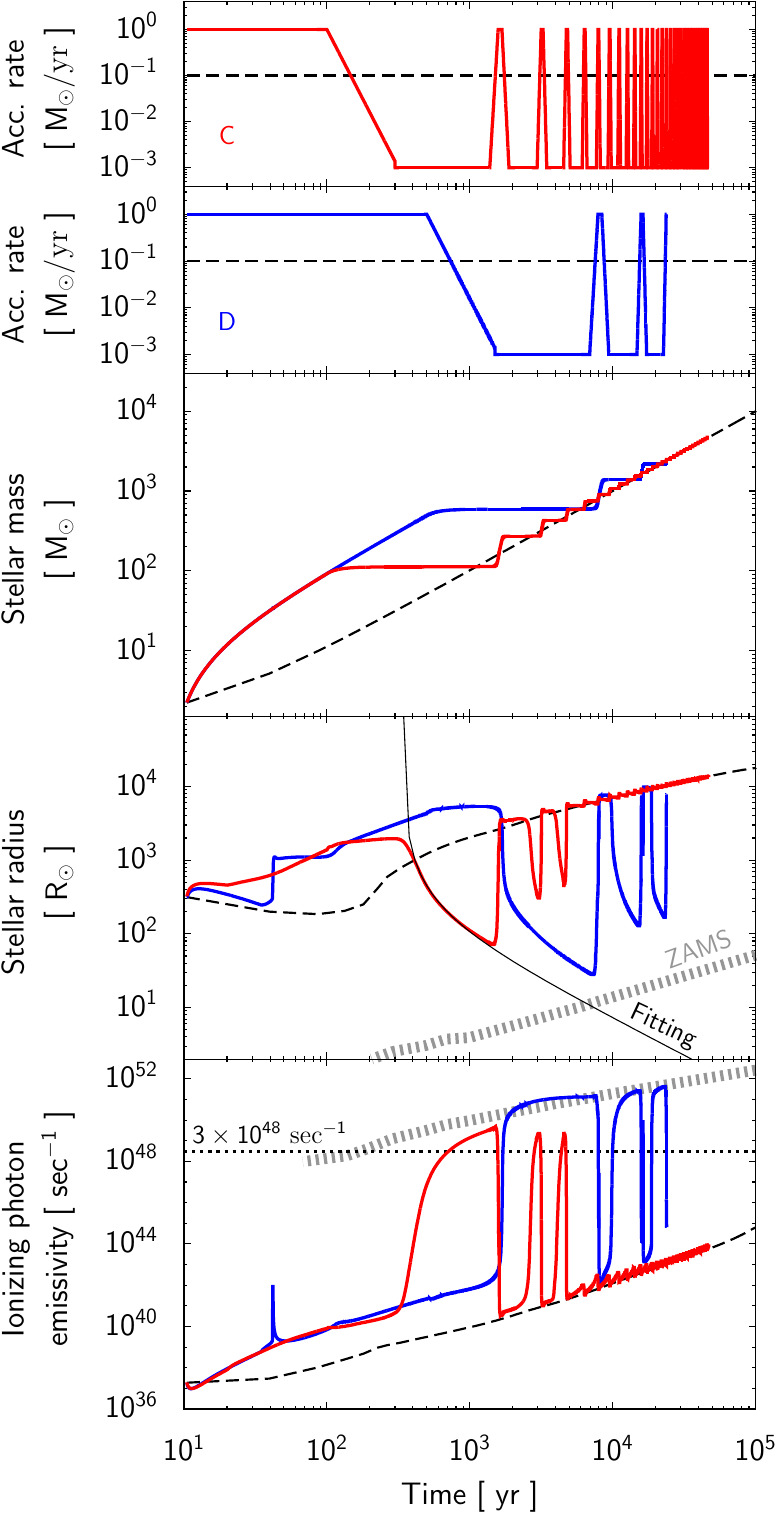}}
 \caption{
 The same as \figref{fig:MDOTMR} but for Models C 
 and D (red and blue lines respectively, see \tabref{tab:accretion}).
 In the fourth panel,  the black solid line represents a fitting function
 of the KH contraction for Model C (see text and footnote \ref{foot:derivation}).
}
 \label{fig:MDOTMR3}
\end{figure}

\subsection{Evolution with burst accretion}
\label{sec:burst}

\subsubsection{Model C}

We first focus on our Model C as a fiducial case for
stellar evolution with burst accretion. 
In this case, the burst and quiescent phases
have the accretion rates $\dot{M}_{\rm *,b} = 1~\msunyr$ and 
$\dot{M}_{\rm *,q} = 10^{-3}~\msunyr$ for the durations of 
$\Delta t_{\rm b} = 100$ yr and $\Delta t_{\rm q} = 1080$ yr
(Table 1).
The protostellar evolution is shown in
\figref{fig:MDOTMR3}, where we note
substantial differences in the
evolution from the case with the constant rate $0.1~\msunyr$.
First of all, the star contracts during the quiescent phases.
At $300 \lesssim t~({\rm yr}) \lesssim 1500$, for instance, 
the stellar radius decreases to $100~R_\odot$, 
which is more than 10 times smaller than
the super-giant protostar for the case $\dot{M}=0.1~\msunyr$ 
at the same mass.
When the star contracts,
the ionizing photon emissivity increases and 
becomes comparable to that of the corresponding 
ZAMS star.


Because the accretion rate in the quiescent phase is lower 
than the critical rate for maintaining bloating of the star, 
$\simeq 4 \times 10^{-2}~\msunyr$,
the star contracts as in the normal KH contraction stage.
To see this, we analytically derive an equation which describes
the time evolution of the radius assuming that the contraction occurs
over the KH timescale,
\begin{equation}
\frac{R_*}{R_\odot}=\left(\frac{1}{R_{*,\mathrm{0}}/R_\odot}
+C\frac{t-t_\mathrm{0}}{1~\mathrm{yr}}\frac{1}{M_*/M_\odot}\right)^{-1}~,
\label{eq:contraction}
\end{equation}
where $C$ is a parameter, and 
the subscript 0 represents the quantities when the contraction 
begins\footnote{\label{foot:derivation}
Derivation of Eq. \eqref{eq:contraction} is as follows.
Assuming that the stellar radius decreases roughly exponentially with
a timescale given by Eq. \eqref{eq:KH}, i.e. the KH time, 
the time derivative of the stellar radius is
\begin{equation}
\ddt{R_*}{t}\sim-\frac{R_*}{t_\mathrm{KH}}=-\frac{L_*R_*^2}{GM_*^2}~.
\label{eq:ddtR/t}
\end{equation}
Using the fact that stellar luminosity is well approximated by the Eddington 
value $\propto M_*$, Eq. \eqref{eq:ddtR/t} becomes
\begin{equation}
\ddt{R_*}{t}=-\mathrm{const.}\times\frac{R_*^2}{M_*}~.
\label{eq:drdt}
\end{equation}
Because the stellar mass is almost constant during a quiescent phase, 
integrating Eq. \eqref{eq:drdt} gives Eq. \eqref{eq:contraction}.
}. 
Figure~\ref{fig:MDOTMR3} shows the fitting curve using 
Eq.~(\ref{eq:contraction}) together
with our numerical result.


The protostar expands again when the next burst accretion occurs
and the ionizing photon emissivity 
decreases accordingly. This cycle is repeated for the following two cycles
of quiescent and burst phases for
$2000 \lesssim t~({\rm yr}) \lesssim 5000$. 
As the star becomes more massive, the features of the contraction gradually diminish.
The star contracts very little during quiescent phases
when $t \gtrsim 5000$ yr, i.e., after the stellar mass
exceeds $500~\msun$.
Thus, UV feedback can become effective with this
burst accretion model, but only for early evolutionary times.

\begin{figure}
 \centering
\resizebox{80mm}{!}{ \includegraphics{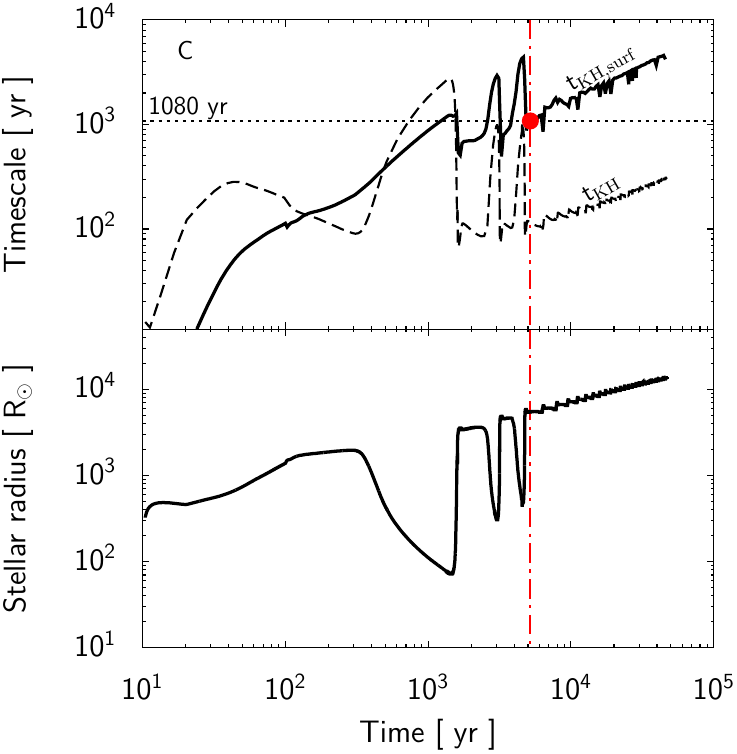}}
 \caption{
 {\it Upper panel:}
 We compare the KH timescales with
 the duration of the quiescent phase $\Delta t_{\rm q}$ 
 (horizontal short dashed line) in Model C.
 The solid and dashed lines show the time evolution of the local
 KH timescale for the bloated surface layer $t_\mathrm{KH,surf}$
 (Eq.~\ref{eq:KHsurf}) and the normal KH timescale 
 $t_\mathrm{KH}$ (Eq.~\ref{eq:KH}).
 {\it Lower panel:} the evolution of the stellar radius in Model C.
 The red circle in the upper panel and vertical dot-dashed line 
 mark the point where $\Delta t_{\rm q} = t_{\rm KH,surf}$.
Thus, the star can appreciably contract during 
 quiescent phases only for earlier times.
 }
 \label{fig:KHtime3}
\end{figure}

The result is understood by comparing the duration of the quiescent
phase $\Delta t_{\rm q}$ with the surface KH time $t_{\rm KH, surf}$
defined in Eq.~\eqref{eq:KHsurf}. 
For the integration range in Eq.~\eqref{eq:KHsurf}, we 
consider a layer covering $0.01~R_* \leq r \leq R_*$ that 
has only $\simeq 10-30$~\% of the total mass
but 99.9999\% of the volume.
As seen in \figref{fig:KHtime3}, the surface KH time 
monotonically increases
with time (and stellar mass), except for some
spiky features that appear when the protostar contracts. 
The mass-radius relation deviates from Eq. (\ref{eq:ranalytic})
for a brief time period.
Note that Eq.~\eqref{eq:KHsurf}
can also be written as 
\begin{equation}
t_\mathrm{KH,surf} \sim
\frac{ f\int Gm/r \mathrm{d}m }{\int \mathrm{d}l}
\propto \frac{GM_*^2}{L_*R_*}\propto M_*^{1/2}~.
\label{eq:KHsurfpropto}
\end{equation}
The first relation comes from the hydrostatic equilibrium 
for radiation pressure, $Ts_{\rm rad}\sim P_{\rm rad}/\rho \sim Gm/r$.
The KH time increases because the
stellar gravitational energy increases with mass as 
$\propto M_*^2/R_* \propto M_*^{3/2}$ whereas the stellar 
luminosity only increases as $\propto M_*$.


\figref{fig:KHtime3} shows that, for 
$t \lesssim 5 \times 10^3$ yr, the surface KH time is 
shorter than the duration of the quiescent phases. 
In other words, the quiescent phase is long enough for the protostar
to lose the thermal energy trapped in the surface
layer. The protostar then contracts.
Even in such cases, however, the stellar contraction
does not immediately follow the decrease of the accretion rate.
After the first burst, for instance, the accretion rate 
falls down $\dot{M}_{\rm cr}$
at $t \simeq 150$ yr (Fig.~\ref{fig:MDOTMR3}),
but the star begins to contract only at $t \simeq 400$ yr.
\figref{fig:KHtime3} shows that the surface KH time 
in this epoch is really $\sim 100$ yr, which explains the
above delay of the contraction.
\figref{fig:MDOTMR3} shows that this delay lengthens
to $\sim 10^3$ yr after the second and third bursts,
because the surface KH time increases with
increasing the stellar mass as shown in \figref{fig:KHtime3}.


For $t \gtrsim 5 \times 10^3$ yr, the surface KH time 
is longer than the duration of the quiescent phases.
The quiescent phase is now too short for the star 
to significantly contract and the next accretion burst begins before 
the stellar surface layer can appreciably relax.
Consequently, the evolution of the star after this point is 
almost the same as in the case with the constant accretion rate $0.1~\msunyr$.

 
\begin{figure}
 \centering
 \includegraphics[width=8cm]{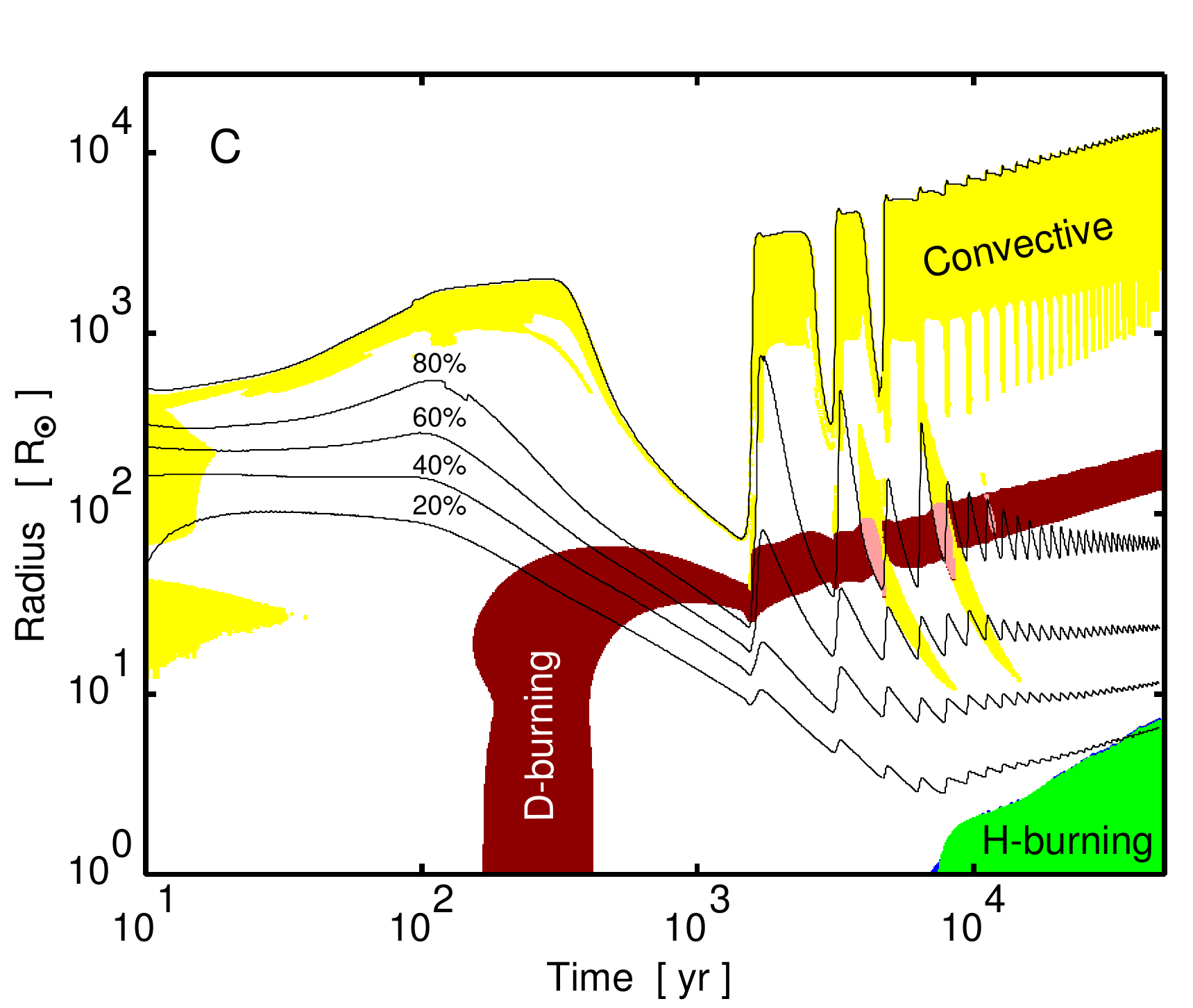}
 \caption{
 The evolution of the stellar interior structure in Model C. 
 The black solid lines indicate the positions of the mass coordinates
 for 100\% (stellar surface), 80\%, 60\%, 40\% and 20\% of the 
 total stellar mass in descending order.   
 The yellow regions denote layers with convection, according to
 mixing length theory. 
 The brown regions show deuterium-burning layers, where the 
 timescale over which deuterium mass fraction decreases by
 nuclear fusion is shorter than the main-sequence lifetime. 
 The green region represents a hydrogen-burning convective core, 
 which is identified measuring the depletion time for hydrogen.
 The pink regions are deuterium-burning convective layers.
 The stellar mass is roughly equal to $\overline{\dot{M}}t$ for 
 $t\gtrsim10^3~\mathrm{yr}$, 
 where $\overline{\dot{M}}$ is the mean accretion rate and $t$ is time.
 }
 \label{fig:structure}
\end{figure}

\begin{figure}
 \centering
\resizebox{80mm}{!}{ \includegraphics{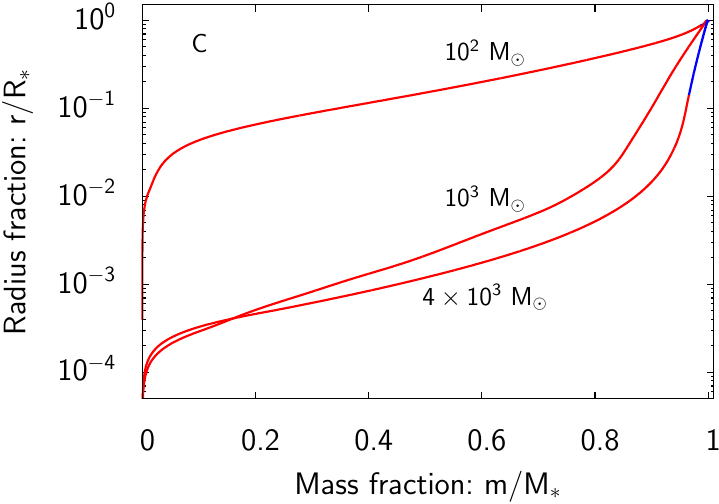}}
 \caption{
The mass distributions in the stellar interior in Model C.
The three snapshots for $M_* = 10^2~M_\odot$ ($t = 10^2$~yr),
$10^3~M_\odot$ ($10^4$~yr) and $4 \times 10^3~M_\odot$ 
($4 \times 10^4$~yr) are presented. The latter two snapshots are 
for the largely bloated protostars. Each profile is plotted against 
the normalized mass coordinate $m/M_*$, where $m$ is the normal mass 
coordinate and $M_*$ is the stellar mass.
The vertical axis also represents the normalized radial distance
from the center, $r/R_*$.
 }
 \label{fig:MRburst}
\end{figure}


As discussed in \secref{sec:0.1}, it is important to use $t_\mathrm{KH,surf}$
as the thermal relaxation time for the bloated protostars instead of
the normal KH time $t_\mathrm{KH}$ since the latter does not consider the 
inhomogeneous density structure in the stellar interior. 
(see Figs. \ref{fig:structure} and \ref{fig:MRburst}). 


Since the normal KH time represents the thermal relaxation time
for the whole star, it might appear puzzling that $t_{\rm KH}$ is shorter
than the local surface KH time $t_{\rm KH, surf}$ for only a part of the star. 
The normal KH time provides the thermal relaxation time if the
following condition is satisfied: 
\begin{equation}
\int_0^{M_*} \frac{Gm}r \mathrm{d}m \sim \frac{G M_*}{R_*},
\end{equation}
which is not true in our case. Since most of the stellar
mass is concentrated near the center as shown in 
Figures~\ref{fig:structure} and \ref{fig:MRburst}, we have
\begin{equation}
\int_0^{M_*} \frac{Gm}r \mathrm{d}m \sim 10 - 100 ~ \frac{G M_*}{R_*},
\end{equation}
depending on the stellar mass and accretion history.
As a result, the effective thermal relaxation time for the whole star is now
$t_{\rm KH, eff} \sim 10-100 \times t_{\rm KH}$, which is always larger 
than the surface KH time $t_{\rm KH, surf}$.

We confirmed that $t_{\rm KH, eff}$ becomes comparable to $\Delta t_{\rm q}$
at $t \sim 1,000$ yr, which does not match the epoch when
the star stops contracting. For the bloated protostar, only entropy
near the stellar surface determines whether the star contracts or not.
Therefore it is reasonable to use the local KH time $t_\mathrm{KH,surf}$
for the surface layer only.

\subsubsection{Variations with different burst accretion}
\label{sec:variation}

\begin{figure}
 \centering
\resizebox{80mm}{!}{\includegraphics{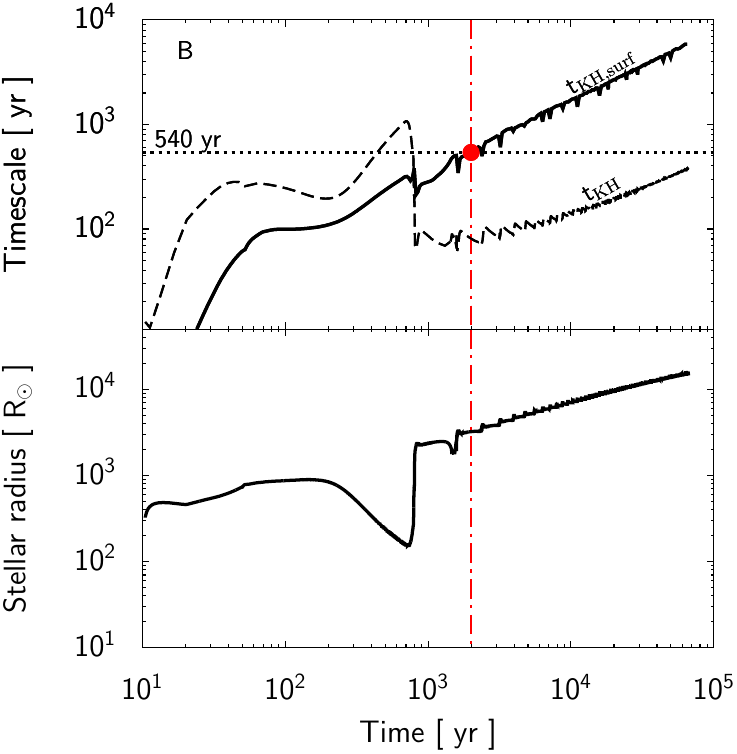}}
 \caption{
 Same as \figref{fig:KHtime3} but for Model B.
 }
 \label{fig:KHtime}
\end{figure}

We can explain the results of Models A, B, and D
based on our findings for Model C. 
In model B, for example, the star does not significantly
contract during quiescent phases for $t \gtrsim 2,000~\mathrm{yr}$
(Fig.~\ref{fig:MDOTMR}).
\figref{fig:KHtime} shows the evolution of the KH time for Model B.
We see that $t_\mathrm{KH,surf}$ becomes equal to the duration of
the quiescent phase $\Delta t_{\rm q}$ at $t \simeq 2,000~{\rm yr}$,
which agrees with the numerical results.
   
Figures~\ref{fig:MDOTMR} and \ref{fig:MDOTMR3} show that, 
as expected, the star stops contracting earlier for
quiescent phases of shorter duration.
Unless the star contracts for a long time, the resulting increase
of the stellar ionizing photon emissivity is not large.
We conclude that the episodic accretion causes the stellar
contraction and potential UV feedback only if the quiescent phase
lasts for $\gtrsim 10^3$ yr.

\section{Discussion}
\label{sec:discussion}

\subsection{UV feedback from supermassive stars evolving with burst accretion}
\label{sec:UV feedback}

We first discuss briefly whether UV photons from a SMS 
growing in mass due to burst accretion can
ionize the surrounding gas to initiate strong radiative feedback.
At an accretion rate $\dot{M}=0.1~\msunyr$, 
the total number of hydrogen and helium atoms infalling
per second is $3 \times 10^{48}~\mathrm{sec^{-1}}$.
Assuming that the feedback becomes effective if all the atoms
are ionized, we derive an approximate lower limit of ionizing 
photon emissivity which induces strong feedback:
$S_\mathrm{min} = 3\times 10^{48}~\mathrm{sec^{-1}}$.
This critical value is shown by the dashed horizontal lines in
the bottom frames of
Figures~\ref{fig:MDOTMR} and \ref{fig:MDOTMR3}. 
In Models A and B, for which the duration of the quiescent phase
$\Delta t_{\rm q}$ is short, the ionizing photon emissivity 
is always lower than $S_\mathrm{min}$ (\figref{fig:MDOTMR}).
By contrast, in Models C and D with the longer quiescent phase
duration $\Delta t_{\rm q}$, the stellar ionizing photon emissivity exceeds
$S_\mathrm{min}$ when the star contracts in the quiescent phases.
Radiative UV feedback becomes effective in quiescent
phases when $\Delta t_{\rm q} \gtrsim 10^3$ yr, 
even if the mean accretion rate is $0.1~\msunyr$.


The above estimate provides only a necessary condition for appreciable
UV feedback, because UV photons are also consumed
by re-ionizing the recombining gas within an \hii region.
Moreover, even if an \hii region appears around the star, 
its expansion is interrupted 
by the burst accretion.
Suppose that the mass supply from the envelope to the
disk continues after an
\hii region is formed. This is likely because the \hii region first
extends only in polar directions with respect 
to the disk in an early
evolutionary stage \citep[e.g.,][]{HOYY11}.
With its mass increasing, the disk becomes gravitationally unstable,
which then can cause another event of burst accretion. 
The stellar contraction is halted and
the ionizing photon emissivity suddenly decreases as the star 
bloats again (see Figs.~\ref{fig:MDOTMR} and \ref{fig:MDOTMR3}). 
As a result, the \hii region quickly disappears. 
Some of the gas expelled by the expanding \hii region might 
fall back onto the disk before the
\hii region reappears during the next stellar contraction.
Thus, it is still uncertain to what extent the star can continue to grow
within the context of burst accretion and intermittent UV feedback.
Further studies are clearly needed to examine the overall
impact of burst accretion on disk accretion and \hii region formation.

\subsection{Accretion histories in atomic-cooling halos}

In this paper, we have modeled the time-dependent accretion
histories using a simple functional form with several free
parameters (Table 1). 
Despite progresses in 3D numerical simulations, 
there remains significant uncertainty in
the long-term accretion history of a supermassive star.
\citet{Latif13b} follow the long-term evolution in
the protostellar accretion stage using the so-called sink
cell technique. Their obtained accretion history shows
some time-variability, but is overall rather smoother 
than considered in our paper. 
\citet{Regan:2014rf} and \citet{Becerra14} use simulations
with a much higher spatial resolution and find
more signatures of the disk fragmentation. 
Although their simulations only follow the initial 
$10 - 100$ yr in the accretion stage, the results 
suggest that a highly time-dependent mass accretion can
be realized in the direct collapse model. 


\citet{Vorobyov:2013lr} perform high resolution 2D simulations
that follow the long-term evolution of a disk for normal Pop III cases.
Episodic accretion occurs in these simulations
with the mean accretion rate of $\sim 10^{-3}~\msunyr$.
The mean value represents the rate onto a disk from the surrounding envelope.
When gas accretion occurs through a self-gravitating disk, 
the accretion rate onto a star becomes strongly variable. 
When fragmentation occurs in the disk, clumps migrate inward 
to the star, and the accretion rate rises up to $\sim 10^{-2} - 0.1~\msunyr$.
In the intervals between such burst events, the accretion rate falls
to $\sim 10^{-5} - 10^{-4}~\msunyr$.


We also expect burst accretion with large variations around the mean 
value of $\sim 0.1~\msunyr$ in the direct collapse model.
If the accretion rates during the quiescent phase 
falls to values $\dot{M}_{\rm *,q} \lesssim 10^{-2}~\msunyr$,
the star can contract as suggested by our results. 


Our calculations show that the duration of the quiescent phase is 
a key parameter that determines if the star contracts or not.
The quiescent phase will be controlled by the following two timescales, 
the effective fragmentation time $t_\mathrm{frag}$ (i.e. taking into account
that multiple clumps can be produced during the fragmentation process)
and the migration time $t_\mathrm{mig}$.
The former is the average time for a fragment to be born in
a gravitationally unstable disk or 1/(rate of clump formation).
The latter is the timescale over which the newly formed fragments migrate
inward to reach the star.
We expect $\Delta t_{\rm q} \sim t_\mathrm{frag} + t_\mathrm{mig}$. 
If $t_\mathrm{frag} < t_\mathrm{mig}$, the duration of the quiescent phase
will be mainly determined by $t_\mathrm{mig}$.
Conversely, if $t_\mathrm{frag}>t_\mathrm{mig}$, the duration is limited by 
$t_\mathrm{frag}$.


Previous numerical and analytical studies give reasonable estimates 
for the timescales. For example, 
\citet{Vorobyov:2013aa} argue that, based on their 
simulations of present-day star formation, 
the fragmentation time should be determined by the timescale 
over which the disk gets mass supply from the accretion envelope, 
\begin{equation}
t_\mathrm{frag}=\frac{M_\mathrm{d}}{\dot{M}_\mathrm{d}}~,
\end{equation}
where $M_\mathrm{d}$ is the disk mass and $\dot{M_\mathrm{d}}$ is the 
infall rate onto the disk.


\citet{Inayoshi:2014qy} develop an analytic model of the structure 
of an accretion disk around a SMS, and estimate where
gravitational fragmentation occurs in the disk.
According to their model, there is a maximum radius 
$R_{\rm f}$, within which the disk is unstable to fragmentation.
The migration time for a fragment formed at $R_{\rm f}$ 
is then estimated to be
\begin{equation}
t_\mathrm{mig,max}\simeq4\times10^3~\mathrm{yr}
\end{equation}
for the mean accretion rate $0.1~\msunyr$
and the central star with $M_* \lesssim 10^4~\msun$.
This can be considered as the {\it maximum} timescale because
a fragment formed at $R<R_{\rm f}$ will have the shorter migration
time, $t_\mathrm{mig} < t_\mathrm{mig,max}$.


We summarize the evolution of $\Delta t_{\rm q}$ expected from the
above discussion as follows. 
In an early evolutionary phase for $t_\mathrm{frag} < t_\mathrm{mig}$,
the quiescent period $\Delta t_{\rm q}$ is limited by 
the migration time $t_{\rm mig}$.
In this case, the quiescent phase lasts for
$10^3~\mathrm{yr}$, which is the critical value for the formation
of an \hii region (Sec.~\ref{sec:variation}), or even shorter.
With the accretion rate $0.1~\msunyr$ from the envelope onto the
disk, however, the fragmentation time $t_{\rm frag}$ exceeds the migration 
time when the disk mass reaches $\simeq 400~\msun$.
After this point, the quiescent phase may become longer 
with increasing the stellar mass, if the disk mass
is comparable to the stellar mass.
Unlike in the case for constant duration $\Delta t_{\rm q}$ studied in
the present paper, the star with high mass will further contract 
during prolonged quiescent phases.
The stellar ionizing photon emissivity should be enhanced by this effect.


Ultimately, in order to verify the above expectation, 
we need to derive
realistic accretion histories realized in an atomic-cooling halo. 
To this end, multi-dimensional hydrodynamic simulations would be
necessary that follow the dynamic accretion process onto 
growing SMSs. Future studies combining radiation hydrodynamics 
simulations and
stellar evolution calculations should yield more realistic models
\citep[e.g.,][]{HOYY11, Rowan12, Machida13, Kuiper13, Hirano14}.

\section{Conclusions}
\label{sec:Conclusions}

We have studied the evolution of supergiant protostars 
growing under episodic accretion, 
which is expected in a self-gravitating
circumstellar disk forming in an atomic-cooling halo
\citep[e.g.,][]{Vorobyov:2013lr,Regan:2014rf,Inayoshi:2014qy}.
We have  followed the protostellar evolution with various assumed
accretion histories of repeating short bursts and long quiescent phases.
In order to examine potential impact of the burst accretion model, 
the adopted accretion histories have the same mean accretion rate
$0.1~\msunyr$ but different variabilities (Table 1).
Our findings are summarized as follows:


Burst accretion can qualitatively change 
the evolution of an accreting SMS. 
Although our previous calculations show that the stellar radius
monotonically increases with increasing the mass under the
constant accretion (Eq.\ref{eq:ranalytic}), this is not always the case
within the framework of burst accretion; i.e. the star can contract during the
quiescent phases between the bursts.
During the contraction, the stellar ionizing photon emissivity 
greatly increases because the effective temperature rapidly rises.
For a quiescent phase of length $\Delta t_{\rm q} \gtrsim 10^3$ yr,
the ionizing photon emissivity exceeds $3 \times 10^{48}~\mathrm{sec}^{-1}$, which
allows the formation of an \hii region. As a result, UV feedback 
might hinder the mass accretion onto the star. 


With a fixed duration of the quiescent phase $\Delta t_{\rm q}$,
stellar contraction can occur in an early epoch; for instance,
for $M_* \lesssim 10^3~\msun$ with $\Delta t_{\rm q} \simeq 10^3$ yr.
With a longer $\Delta t_{\rm q}$, however, stellar contraction
will continue until the star becomes much more massive.
This is well understood by comparing the duration of the quiescent phase 
and the surface thermal relaxation (or KH) time $t_{\rm KH, surf}$ 
defined by Eq.~\eqref{eq:KHsurf}. 
As the surface KH time increases with increasing the stellar mass,
the timescale balance changes from $\Delta t_{\rm q} > t_{\rm KH, surf}$
to the opposite at some point. 
Before this, the star contracts significantly during the quiescent phases because
the bloated surface layer loses most of its entropy before the next burst
begins. After the timescale inversion into 
$\Delta t_{\rm q} < t_{\rm KH, surf}$, the star
continues to expand following roughly the mass-radius relation
Eq.~\eqref{eq:ranalytic}. Here, the same comparison using the global 
KH time $t_{\rm KH}$ instead of $t_{\rm KH,surf}$ largely overestimates 
the critical stellar mass above which the stellar contraction ceases. 
This is because the global KH time does not consider the 
inhomogeneous density structure in the stellar interior.
Since the mass distribution is very centrally condensed when the
star is in the supergiant stage (Figs ~\ref{fig:structure} 
and \ref{fig:MRburst}), the global KH time provides a poor estimate of the 
thermal relaxation time. 


Our calculations show that 
stellar radiative feedback may become important, 
if the quiescent phase is longer than $\sim 10^3$ yr.
Such long intervals between bursts of accretion are 
expected for an accretion disk around a growing 
SMS \citep{Inayoshi:2014qy}.
If the stellar mass growth is halted by UV feedback, the
final mass of the SMS is reduced and the mass of the 
remnant BH left behind after the star's death is also reduced.
Although we have adopted a simple functional form with parameters
(Table 1) to model the accretion histories, hydrodynamical
simulations will provide more realistic estimates of the final mass.
The stellar evolution and strength of the resulting UV radiative
feedback will be investigated in future studies.

We thank Kazuyuki Omukai, Kohei Inayoshi and Eduard Vorobyov 
for fruitful discussions.
The calculations were in part carried out on PC cluster at Center for Computational Astrophysics, National Astronomical Observatory of Japan.
This work is supported in part by Grant-in-Aid for
Scientific Research from the JSPS Promotion of Science
(25800102, 15H00776, 25287050, and 15J08816). 
Portions of this research were conducted at the Jet Propulsion Laboratory, 
California Institute of Technology, operating under a contract with the 
National Aeronautics and Space Administration (NASA).


\bibliography{SMS_accvar}


\end{document}